\documentclass{PoS}

\title{Classification of Fermi Gamma-RAY Bursts}

\ShortTitle{Classification of Fermi GRBs}

\author{Istvan Horvath\\
        Bolyai Military University, Budapest, Hungary\\
        E-mail: \email{horvath.istvan@uni-nke.hu}

        }

\author{Lajos G. Bal\'azs\\
        MTA CSFK Konkoly Observatory, Budapest, Hungary\\
        E\"otv\"os University, Budapest, Hungary\\
        E-mail: \email{balazs@konkoly.hu}
        }
        
\author{Jon Hakkila\\
        College of Charleston, Charleston, SC USA\\
        E-mail: \email{hakkilaj@cofc.edu}
        }

\author{Zsolt Bagoly\\
        E\"otv\"os University, Budapest, Hungary\\
        Bolyai Military University, Budapest, Hungary\\
        E-mail: \email{zsolt.bagoly@elte.hu}
        }

\author{Robert D. Preece\\
        University of Alabama in Huntsville, Huntsville, AL USA \\
        E-mail: \email{rob.preece@nasa.gov}
        }

\abstract{
The Fermi GBM Catalog has been recently published. 
Previous classification analyses of the BATSE, RHESSI, 
BeppoSAX, and Swift databases found three types of 
gamma-ray bursts. Now we analyzed the GBM catalog 
to classify the GRBs. PCA and Multiclustering analysis 
revealed three groups. Validation of these groups, 
in terms of the observed variables, shows that one 
of the groups coincides with the short GRBs. 
The other two groups split the long class into a 
bright and dim part, as defined by the peak flux. 
Additional analysis is needed to determine
whether this splitting 
is only a mathematical byproduct of the analysis 
or has some real physical meaning. 
}

\FullConference{Gamma-Ray Bursts 2012 Conference -GRB2012,\\
		May 07-11, 2012\\
		Munich, Germany}

\begin{document}

\section{Introduction}

Accumulating evidence indicates that GRBs represent a mixture 
of objects of different physical natures. Classification of 
GRBs in the parameter space of their observable properties 
may give information on the number and physical nature of 
types of these objects. Mazets et al. \cite{maz81} and Norris et al. 
\cite{nor84} suggest that there might be a separation in the duration 
distribution. In the First BATSE Catalog, a bimodality was found 
in the logarithmic duration distribution \cite{kou}.  
Today it is widely accepted that the physics of these two 
groups ({\em long} and {\em short} GRBs) are different, 
and these two kinds of GRBs represent different 
phenomena (\cite{bal03}, 
\cite{fox05}, \cite{nor01},  \cite{zha07}). Zhang et al. \cite{zha09} uses 
Type I. and II. classification based on the progenitor models. 
		
In a previous paper using the Third BATSE Catalog 
Horváth \cite{ho98} has shown that the duration ($T_{90}$) 
distribution of GRBs observed by BATSE could be well fitted 
by a sum of three log-normal distributions. 
Simultaneously, Mukherjee et al. \cite{muk98} reported the 
finding (in a multidimensional parameter space) of a 
very similar group structure of GRBs.  Somewhat later 
several authors (\cite{bala}, \cite{chat07}, \cite{hak}, 
\cite{hak03}, \cite{lu10}, \cite{rm}, \cite{deu11}) included more 
physical parameters into the analysis of the bursts 
(e.g. peak-fluxes, fluences, hardness ratios, etc.). 
The physical existence of the third 
group is, however, still not convincingly proven. However, 
the celestial distribution of the third group is anisotropic 
(\cite{li01},  \cite{mgc03},  \cite{me00b},  \cite{vav08}).  
All these results mean that the 
existence of the third {\em intermediate} group in the BATSE  \cite{ho06}, 
RHESSI  \cite{ri}, BeppoSAX  \cite{ho09} and Swift  \cite{ho08}  \cite{ho10} sample 
is acceptable, but its physical meaning, importance and origin 
is less clear than those of the other groups. Hence, it is 
worth to study new samples if their size is large enough for 
statistics. In this paper we use the new Fermi catalog 
\cite{pac12} data for this analysis.

\begin{figure}[h!]
\begin{center}
\includegraphics[height=.8\linewidth,angle=0]{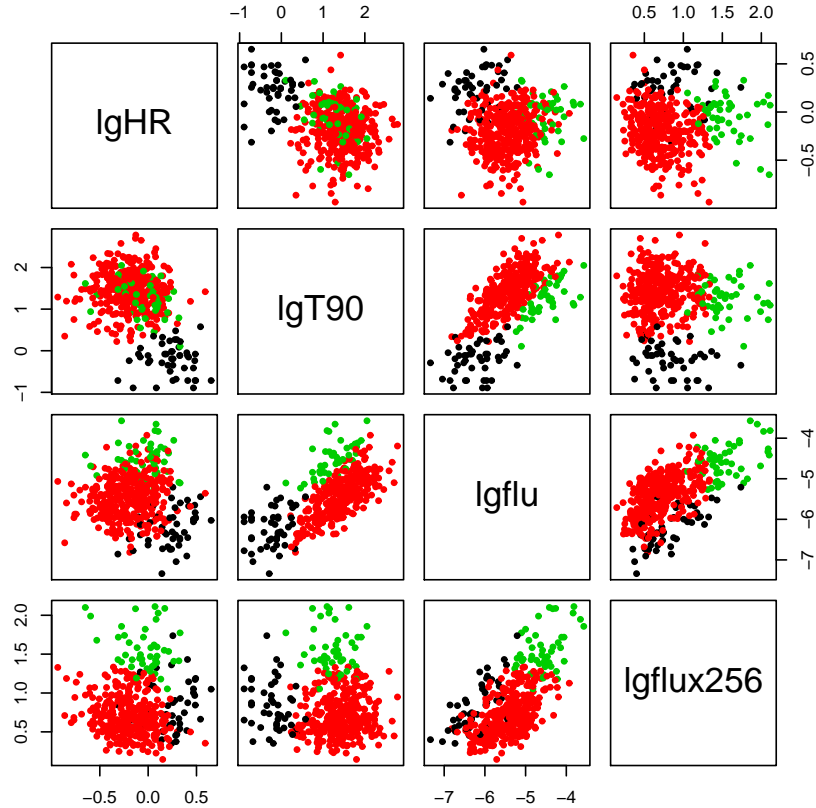}
\end{center}
\caption{The color coded three groups in the Fermi data. 
The black circles are the short, the red and green
ones are the long GRBs. Green circles are the brighter
bursts among the long ones.
}
\label{fig2}
\end{figure}

\section{Analysis of the Fermi GMB data}

The Fermi GRB catalog \cite{pac12}
contains 490 GRBs, we used 425 GRBs which had no huge errors 
in the observed parameters. From the Catalog we used 
the following variables $T_{90}$, total fluence, hardness 
ratio and peakflux256.
Using the correlation matrix we made a Principal Component 
Analysis (PCA).  
The eigenvalues are 1.915, 1.305, 0.725 and 0.055. 
The computations were made with the R statistical package. 

After PCA we used {\em Mclust}, the clustering algorithm fitting of a 
functional model superposed from Gaussian components. 
Optimizing the number of the components and their parameters 
is the task of the procedure. 
Since the first 3 PC represent  98.8 \% of the total 
variance of the Fermi data, respectively, we kept 3 
PCs as input variables for the Mclust procedure. This choice 
is reasonable because the first 3 PCs 
account for the vast majority of the variances of the observed variables, used in 
the analysis (PC1 accounts mainly for lgT90 and lgfluence, 
PC2, PC3 do it for lgHR and lgp256). The best fitting model is a three 
component Gaussian, coded by 'EEE', meaning equal volumes, 
equal shapes and parallel axes of the error ellipses. 
To study the impact of classifications on the input data we 
computed cluster membership of the bursts and used as a color 
code in a matrix plot displaying the dependencies among the 
observed variables (Figure 1.).

\section{Conclusion}

The logarithmic duration, fluence, hardness and peakflux 
variables can be well represented by three PCs obtained from 
PCA of the correlation matrix. The PC1 accounts for the duration 
and fluence while PC2, PC3 do it for the hardness and peakflux. 
The best fit of a model family of superposed multivariate 
Gaussian functions revealed three groups. Validation of 
these groups, in terms of the observed variables, showed 
that one of the groups coincides with the short GRBs. 
The other two ones split the long group into a bright and 
dim part according to the peak flux; a result similar to this has
been found previously for BATSE data \cite{roi}. Further 
analysis is needed to
determine whether this splitting is only a mathematical byproduct 
of the analysis or whether it has some real physical meaning.

\acknowledgments{
This work was supported by the Hungarian OTKA-77795 grant. 
JH acknowledges support from NASA-AISRP NNX09AK60G and NASA-ADAP NNX09AD03G. 
}

\end{document}